\documentclass[11pt,a4paper]{article}
\usepackage{graphicx}
\usepackage{subcaption}
\usepackage{url}
\usepackage{authblk}
\usepackage{lineno} 
\usepackage[T1]{fontenc}
\usepackage{newpxtext,newpxmath}
\usepackage{setspace}

\usepackage{amsmath} 
\usepackage{amssymb}
\usepackage{hyperref} 
\usepackage{booktabs}
\usepackage[table,xcdraw]{xcolor}
\usepackage{caption}
\usepackage{booktabs}
\usepackage{multirow}
\usepackage{rotating} 
\usepackage{hhline}
\usepackage[square,sort,comma,numbers]{natbib}
\usepackage{url}

\usepackage{epstopdf, epsfig}

\title{Psycho-linguistic differences among competing vaccination communities on social media}

\author[1]{Jialiang Shi}
\author[2]{Piyush Ghasiya}
\author[2]{Kazutoshi Sasahara}

\affil[1]{Graduate School of Informatics, Nagoya University, Furo-cho, Chikusa-ku, Nagoya 464-8601, Japan}
\affil[2]{School of Environment and Society, Tokyo Institute of Technology, 3-3-6 Shibaura Minato-ku, Tokyo 108-0023, Japan}
\date{\small Corresponding authors: sasahara.k.aa@m.titech.ac.jp}

\begin{document}
\maketitle

\begin{abstract}
Currently, the significance of social media in disseminating noteworthy information on topics such as health, politics, and the economy is indisputable. During the COVID-19 pandemic, anti-vaxxers have used social media to distribute fake news and anxiety-provoking information about the vaccine. Such social media practice may harm the public. Here, we characterise the psycho-linguistic features of anti-vaxxers on the online social network Twitter. For this, we collected COVID-19 related tweets from February 2020 to June 2021 to analyse vaccination stance, linguistic features, and social network characteristics. Our results demonstrated that, compared to pro-vaxxers, anti-vaxxers tend to have more negative emotions, narrative thinking, and immoral tendencies. 
Furthermore, we found a tighter network structure in anti-vaxxers even after mass vaccination.
\end{abstract} 

\section{Introduction}
The opposition to vaccines or anti-vaccinationism is an old phenomenon with roots traced back to the 1850s. Much of the anti-vaccine sentiment of the era was laid out by John Gibbs, when he published a booklet against the Vaccination Act of 1853 (British Government) --- which required mandatory vaccination for all infants over four months old \cite{day2021jonathan}. There are several reasons for this, such as conspiratorial beliefs, disgust, individualism, and hierarchical worldviews for anti-vaccine behaviour or attitudes of people \cite{hornsey2018psychological}. 

While early anti-vaccine activists distributed pamphlets and organized rallies, the anti-vaxxers of the 21st century have access to Facebook, Twitter, and other social media platforms to spread their views globally. Due to these communication tools, we see extremely potent (small but far reaching) anti-vaccination movements during the current COVID-19 pandemic. Although as of October 17, 2021, 47.5\% of the world's population has received at least one dose of the COVID-19 vaccine \cite{mathieu2021global}, and the immunization program is going strong, researchers believe that the present anti-vaccination movement can undermine the efforts to end the coronavirus pandemic \cite{ball2020anti}. Therefore, at a time when the COVID-19 pandemic has killed millions of people, it is essential to counter those who actively spread vaccine-related mis/dis-information, fake news, conspiracy, and propaganda on social media platforms, to make it possible to immunize all vulnerable people against the deadly COVID-19 disease.

As social media plays an important role in propagating global anti-vaccination beliefs against the COVID-19 vaccine, several researchers are analysing anti-vaxxers behaviour and attitude on social media \cite{keelan2010analysis}. However, few researchers have tackled questions related to the psychological and moral aspects in the language of anti-vaxxers on social media. By understanding the social media posts of anti-vaxxers from psychological, moral, and linguistic perspectives, we can peek into their minds, which can be helpful in developing counter strategies against the anti-vaccination movement, both at individual and platform levels \cite{burki2020online}. 

In this research, we used longitudinal data that captures COVID-19 discussions on Twitter. First, we utilized the Louvain algorithm to distinguish between anti-vaxxers and pro-vaxxers. Then we used the Linguistic Inquiry and Word Count (LIWC) 2015~\cite{pennebaker2015development} to quantify their semantic differences, and then used the Moral Foundation Dictionary (MFD)~\cite{graham2009liberals} to compare moral tendencies in pro- and anti-vaxxers.
As explained later, these are powerful tools to capture anti-vaxxers' characteristics of spontaneous social media posts implicated in linguistic information.
We found that as the COVID-19 pandemic spread and vaccination started, the psycho-linguistic features of anti-vaxxers' posts also changed. 

Here we pose two research questions (RQs):
\begin{itemize}
    \item \textbf{RQ 1}: How do anti-vaxxers differ from pro-vaxxers in terms of psycho-linguistic features as well as moral values they convey?
    \item \textbf{RQ 2}: How did the mass vaccination (i.e. December 2020) affect pro- and anti-vaxxers' behaviour? 
\end{itemize}
Regarding RQ1, previous research has reported that anti-vaxxers use analytical language less frequently~\cite{memon20201characterizing} and specific moral foundations are associated with vaccine hesitancy~\cite{amin2017association,heine2021using}. 
We intend to confirm if these psycho-linguistic properties are observed in our data. 
We can leverage such linguistic signals to better detect harmful content from anti-vaxxers.
Regarding RQ2, various studies have analysed public moods using social data, but emotional changes have not been measured before and after mass vaccination concerning the evolution of anti-vaxxers.
Therefore, we measured mass vaccination effects on emotions in pro- and ant-vaccine groups to get hints on vaccine operations. 

Answering these questions and understanding the whole picture of anti-vaxxers helps us develop countermeasures against the online anti-vaccine movement. As detailed later, our results indicate that we need to pay attention to the apparent negative tendency of anti-vaxxers in language expressions and the fact that they are resolute in their beliefs in network structure.

\section{Related Work}
\label{sec:related_work}
Various social media platforms work as conduits in the circulation and amplification of fake news \cite{kalsnes2018fake}. Health-related misinformation and fake news is also a burgeoning research topic among social scientists and medical research professionals. In this line of research, a recent study by Suarez-Lledo et al. investigated health misinformation on social media and found that `vaccine' is the fastest spreading topic on Twitter \cite{suarez2021prevalence}.
YouTube is another platform on which anti-vaccine narratives are often broadcast. Lahouti et al. investigated YouTube videos to understand anti-vaccine sentiments in France where vaccine hesitancy is high and found that anti-vaxxers are very active on YouTube \cite{lahouati2020spread}. 

Different demographic characteristics and living conditions also play an important role in anti-vaccine views/beliefs. Lyu et al. showed these differences in their analysis of Twitter data~\cite{lyu2021social}. Cultural differences also play a critical role in developing pro- or anti-vaccine views. A study by Luo et al. observed that the difference between the topic of anti-vaxxers in China and the United States is caused by cultural distinctions between each country \cite{luo2021exploring}. In India, concerns about anti-vaccination are more likely to stem from health concerns and fear of allergic reactions~\cite{praveen2021analyzing}. 

In addition to differences in content and geography, curiosity to understand what kind of arguments anti-vaxxers give on Facebook has also influenced research. Wawrzuta et al. analysed Polish media fan pages and found that the COVID-19 anti-vaccine movement has new arguments, such as the vaccine not being properly tested. However, the classic argument --- not trusting the government --- also remains popular \cite{wawrzuta2021arguments}. Nuzhath et al. analysed Twitter data to understand the prominent topics of discussion among anti-vaxxers in Bangladesh and found that misinformation, vaccine safety and effectiveness, conspiracy theories, and mistrust in government are some of the main topics \cite{nuzhath2020covid}. 

Jamison et al. also found that both pro- and anti-vaxxers are spreading less reliable information or claims on social media and suggested that while all research focus is on bad actors (anti-vaxxers) to understand the anti-vaccine movement, good actors (pro-vaxxers) also play a role in the spread of the `infodemic' \cite{jamison2020not}.

LIWC is a useful tool in investigating psycho-linguistic features. Mitra et al. utilized LIWC to understand anti-vaccination attitudes in social media and found that anti-vaxxers tend to be influenced by conspiracy theories \cite{mitra2016understanding}. Linguistic differences often accompany network differences. Johnson et al. utilized network analysis to understand the evolution of pro- and anti-vaccine communities \cite{johnson2020online}. One significant finding of their research is related to undecided individuals. Their finding challenges the current thinking that undecided individuals are a passive background population in the battle of `hearts and minds'. Germani and Biller-Andorno show that, compared to pro-vaxxers, anti-vaxxers on Twitter have a high number of influencers and these influencers lead the anti-vaxxers discussion \cite{germani2021anti}. They also showed that before the suspension of his Twitter account, Donald Trump was the main driver of anti-vaccine misinformation on Twitter. Lastly, Menon and Carley characterized COVID-19 misinformation communities on Twitter \cite{memon20201characterizing}. Their analysis suggested that a large majority of misinformed users may be anti-vaxxers. Further, their socio-linguistic analysis also showed that informed users (who spread true information) use more narrative thinking than misinformed users (who spread misinformation). 

Social and behavioural scientists have worked to understand the moral basis of people's judgment for a long time. Voluminous research and a theoretical framework led the foundation of the Moral Foundations Theory (MFT) \cite{graham2013moral,graham2018moral}. MFT works on the assumption that there are five major moral foundations: (1) `Care/Harm', which focuses on not harming others and protecting the vulnerable; (2) `Fairness/Cheating', which assumes equivalent exchange without cheating to be good; (3) `Loyalty/Betrayal', which concerns a collective entity instead of individuals; (4) `Authority/Subversion', which postulates respect for authority, resulting in maintaining the hierarchy; and (5) `Sanctity/Degradation', which involves a feeling of disgust caused by the impure. 
Moral Foundations Theory (MFT) was also used to understand why morality varies across cultures yet still shows similarities and recurrent themes \cite{graham2013moral}. 

Within vaccination hesitancy, past research also shows that core morality will influence people's attitudes toward vaccination \cite{day2014shifting}. For example, `Liberty' is likely implicated in the decision to not vaccinate a child \cite{beard2017no}; endorsement of the foundations of Purity and Liberty are associated with vaccine hesitancy \cite{amin2017association,heine2021using}. Moral Foundations Dictionary (MFD) is often used to quantify and understand the extent to which moral foundations are expressed in a text~\cite{graham2009liberals}. We used the original version of MFD for this study. There are also some candidates for moral foundations like `Liberty/oppression' related to MFT~\cite{Haidt2013-ew}. However, since the original MFD does not include this dimension, we do not discuss it in this study.

Although several aspects of anti-vaccine communities have been reported by a series of studies (such as those previously mentioned), the psycho-linguistic features of anti-vaccine posts that may increase vaccine hesitancy, especially in the context of the COVID-19 pandemic, remain unclear and understudied. This knowledge is critical to reduce vaccine mis/dis-information and achieve herd immunity towards a post-pandemic era. 
Therefore, we investigate psycho-linguistic properties of anti-vaxxers in terms of the above-mentioned research questions.

\section{Data and Methods}
\subsection{Data collection}
To obtain the longitudinal data of social media posts we used the Twitter Search API to collect COVID-19 related tweets, replies, and retweets (RTs) with keywords such as `covid', and `covid-19' from February 20, 2020 to June 30, 2021. In this research, we only focused on English language content. 

We then filtered vaccine-related tweets from the English dataset for our analyses, using the keywords `vaccination', `vaccines', `vaccine', `vaccinated', `vaccination', `vaccineoutside', `vaccinate', `vaccinologist', `vacciner', and `coronavirusvaccine'. In total we collected 11,395,103 retweets, 11,395,103 tweets, 465,037 replies and 3,781,447 unique users.
Our data and code are available online (\url{https://osf.io/FSM23/}).

\subsection{Characterisation of anti- and pro-vax groups}
Previous studies have found that understanding retweet networks and community detection is a useful method to reveal communication patterns among communities \cite{cherepnalkoski2016retweet}. 
Therefore, we followed the same approach to classify pro- and anti-vaxxers on Twitter. 
We first constructed a retweet network from the vaccine-related retweets. 
A retweet network on Twitter can be defined as a directed weighted graph, where nodes and edges represent users and retweet transmissions, respectively. In our research, the direction from one node to another represents a user retweeting another user's post. The weight represents the number of times the user retweeted another user's post. 

We used the Louvain algorithm, a standard algorithm for community detection~\cite{blondel2008fast}, to find clusters including pro- and anti-vaxxers. We then applied the $k$-core decomposition ($k=1$) and retained nodes whose indegree (i.e., the number of retweets by different users) was greater than 20 in order to focus on significantly influential users.
The Forceatlas2 layout~\cite{jacomy2014forceatlas2} was used to visualize the structure of the resulting clusters.
For these processes we used the Gephi software \cite{bastian2009gephi}.
By manually confirming the top 10 popular users in each cluster (i.e., looking at users' profiles, tweets, and retweeted contents), we deciphered which clusters represent pro- and anti-vaxxers, and others.  

\subsection{Network measures}
The retweet network was also quantified using the following standard measures to illuminate its structural features, in addition to the number of nodes (users) and links (unique retweet relations): 
\begin{itemize}
    \item \textbf{Network density:} the ratio of actual connections and potential connections.
    \item \textbf{Clustering coefficient:} measures the degree to which nodes in a network tend to form triangles or `clusters'. 
    \item \textbf{Average distance:} the average minimum number of connections to be crossed from any arbitrary node to any other.
\end{itemize}
For this measurement, we converted our retweet network to undirected graphs because these measures are defined for undirected graphs.

\subsection{Psycho-linguistic analysis with LIWC}
LIWC is a standard tool in social psychology for computerized text analysis~\cite{pennebaker2015development}. 
Given a text, LIWC can quantify emotions, thinking style, and social concerns by counting the dictionary words registered in LIWC. The psychological expressions in a text can also reveal critical aspects. Table \ref{tab:LIWC2015-information} shows all the LIWC categories and subcategories used in this study. 
LIWC was used in many studies, such as sentiment analysis and social relationships, to evaluate the impact of psychological expression \cite{tausczik2010psychological}. 

We used LIWC 2015 to investigate attitudinal and linguistic differences between pro- and anti-vaxxers. LIWC is especially useful for measuring the sentiment and thinking styles in a tweet. According to \cite{jordan2016candidates}, analytical thinking and narrative thinking are often in opposition, which may characterize pro- and anti-vaxxers. Narrative thinking is identifying conceptual categories and organizing them in hierarchical ways~\cite{jordan2016candidates}, which can be linked to the frequent use of pronouns and function words \cite{pennebaker2014small} and the less frequent use of analytic categories in LIWC.  
Thus, we compare LIWC scores for analytic, pronouns, and function words between pro- and anti-vaxxers to determine whether they are analytic thinkers or narrative thinkers.

\begin{table*}[t]
\centering
\caption{LIWC2015 categories and subcategories with example words}
\label{tab:LIWC2015-information}
\begin{tabular}{llcc}
\hline
\multicolumn{1}{c}{\textbf{Category}} & \multicolumn{1}{c}{\textbf{Abbrev}} & \textbf{Examples} & \textbf{\# of words in the category} \\ \hline
\textbf{Summary Language Variables} &  & \multicolumn{1}{l}{} & \multicolumn{1}{l}{} \\
Analytical thinking & Analytic & - & - \\
\textbf{Total function words} & funct & it, to, no, very & 491 \\
\quad Total pronouns & pronoun & I, them, itself & 153 \\
\quad Articles & article & a, an, the & 3 \\
\quad Prepositions & prep & to, with, above & 74 \\
\quad Auxiliary verbs & auxverb & am, will, have & 141 \\
\quad Common Adverbs & adverb & very, really & 140 \\
\quad Conjunctions & conj & and, but, whereas & 43 \\
\quad Negations & negate & no, not, never & 62 \\
\textbf{Psychological Processes} &  &  &  \\
Affective processes & affect & happy, cried & 1393 \\
\quad Positive emotion & posemo & love, nice, sweet & 620 \\
\quad Negative emotion & negemo & hurt, ugly, nasty & 744 \\
Personal concerns &  &  &  \\
\quad Work & work & job, majors, xerox & 444 \\
\quad Leisure & leisure & cook, chat, movie & 296 \\
\quad Home & home & kitchen, landlord & 100 \\
\quad Money & money & audit, cash, owe & 226 \\
\quad Religion & relig & altar, church & 174 \\
\quad Death & death & bury, coffin, kill & 74 \\ \hline
\end{tabular}
\end{table*}

\subsection{Morality analysis with MFD}
As mentioned in \ref{sec:related_work}, Graham et al. developed the moral foundations dictionary (MFD), which consists of 156 words and 168 word stems to quantify the frequency of words referring to virtues and vices associated with each moral foundation \cite{graham2009liberals}. They attempted to understand moral tendencies among liberals and conservatives and found that liberals consistently showed greater endorsement and use of the Care and Fairness foundations compared to the other three foundations, whereas conservatives endorsed and used the five foundations more equally. The Japanese version of MFD was used to reveal that a trade-off between the Fairness and Authority foundations plays a key role in the online communication of Japanese users on Twitter~\cite{matsuo2021appraisal}.
We also used the MFD in this study to measure five moral foundations as additional psycho-linguistic features for pro- and anti-vaccine clusters on Twitter.

\section{Results}

\subsection{Pro/Anti-vaxxers and other communities}
\begin{figure}[t]
\includegraphics[width=\linewidth]{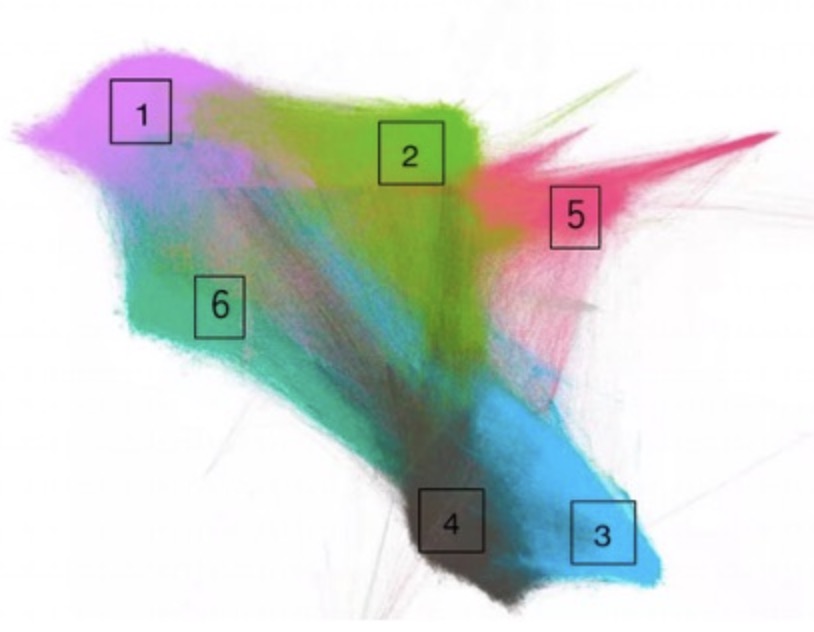}
\caption{Retweet network and communities constructed from vaccine-related tweets: 1. Left-wing (32.0\%) 2. Pro-vaxxers (22.2\%) 3. Right-wing (13.0\%) 4. Anti-vaxxers (11.7\%) 5. India-related (5.7\%) 6. Canada-related (3.6\%).}
\label{fig:cluster-retweet}
\end{figure}

The above-mentioned processes resulted in the retweet network based on vaccine-related tweets (77,934 nodes and 1,999,164 edges). In this network, we identified six main clusters: 1) Left-wing (32.0\%), 2) Pro-vaxxers (22.2\%), 3) Right-wing (13.0\%), 4) Anti-vaxxers (11.7\%), 5) India-related (5.7\%), and 6) Canada-related (3.6\%). These clusters are shown in Fig. \ref{fig:cluster-retweet}. 

We noticed the apparent differences in the content of the top 10 users selected by indegree (i.e., the number of retweets by different users) in each cluster. 
Our examinations identified the second largest cluster as a pro-vaccination group and the fourth largest cluster as an anti-vaccination group. Table \ref{tab:sample of post} shows two example tweets for pro-vaxxers and anti-vaxxers, respectively. Here we can see that anti-vaxxers intimated the Bill Gates conspiracy theory (top) and distrust about the government (bottom).
In addition, we observed that some groups have a clear political orientation, such as the first and third largest groups. In the COVID-19 pandemic, vaccine operation is an important political issue; thus, polemical clusters might emerge together with pro- and anti-vaccine clusters.
Two other groups turn out to be country-related (i.e., India and Canada), which are related to vaccine strategies in these countries.
Because our interest is in psycho-linguistic features in anti-vaccine groups to gain insights into countermeasures, we hereafter restricted our analysis to pro- and anti-vaccine groups (clusters 2 and 4).

\begin{table*}[]
\centering
\caption{Sample tweets of pro- and anti-vaxxers.}
\label{tab:sample of post}
\begin{tabular}{ll}
\toprule
\multicolumn{1}{c}{\textbf{Pro-vaxxers}} & \multicolumn{1}{c}{\textbf{Anti-vaxxers}} \\ \midrule
\begin{tabular}[c]{@{}l@{}}this is a critical step toward \\ developing vaccines and \\ treatments. \#coronavirus \#ncov\end{tabular} & \begin{tabular}[c]{@{}l@{}}\#coronavirus \&amp; the georgia \\ guidestones - what you need to\\  know! \#billgates \#vaccines\end{tabular} \\
 &  \\
\begin{tabular}[c]{@{}l@{}}citizens 60 and above who\\ have not yet been vaccinated\\ can go to their 
nearest\\ vaccination site to be\\ vaccinated. here is a list\\ of vaccination sites:  \# ichoosevaccination\end{tabular} & \begin{tabular}[c]{@{}l@{}}covid-19 vaccine company under\\ federal investigation over allegedly\\ misrepresenting its role\\ in government program\end{tabular} \\ \bottomrule
\end{tabular}
\end{table*}

\begin{figure}[!h]
    \centering
    \includegraphics[width=\linewidth]{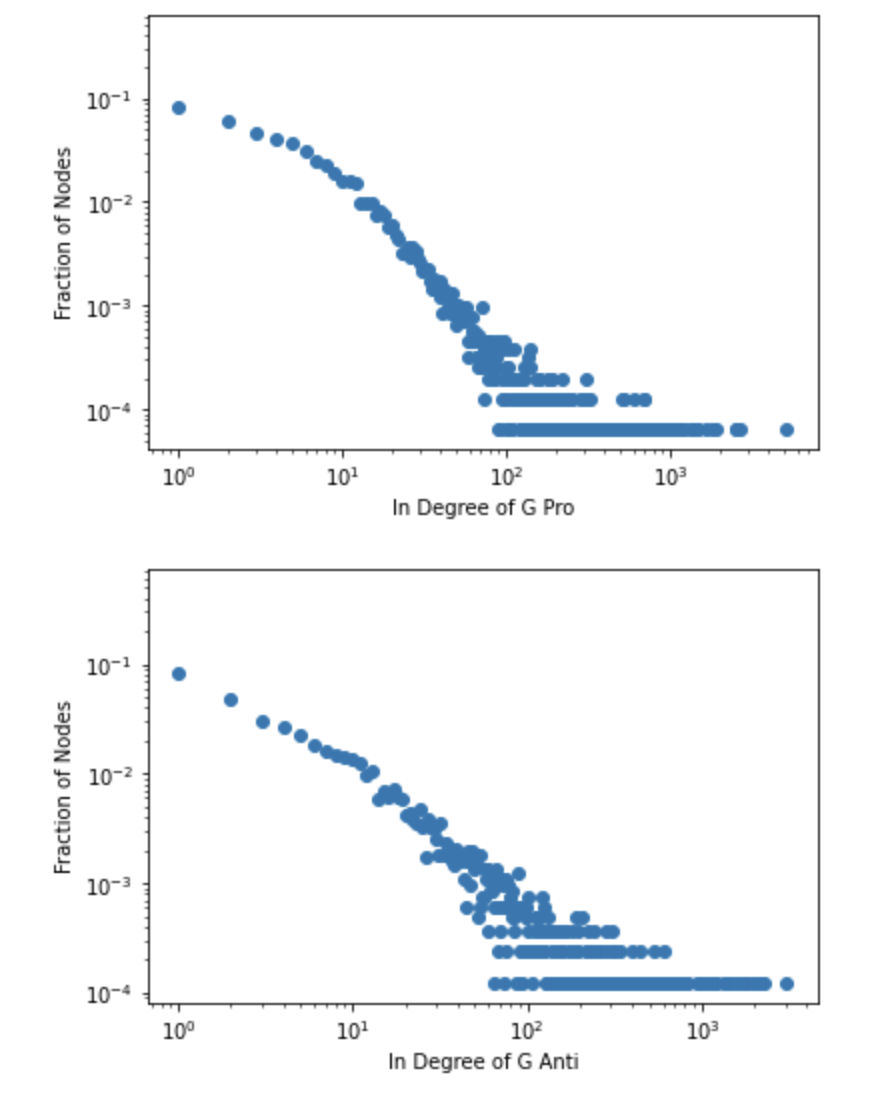}
    \caption{Distribution of the indegree for anti-vaxxers (top) and pro-vaxxers (bottom).}
    \label{fig:Distribution-indegree}
\end{figure}

The network measures for the entire retweet network and the pro- and anti-vax clusters are summarised in Table~\ref{tab:network-analysis}, which reveals that the number of pro-vaxxers is larger than that of anti-vaxxers. However, anti-vaxxers are more densely connected according to network density and clustering coefficient values. In addition, the distributions of the indegree for pro- and anti-vaccine groups have heavy-tailed distributions (Fig.~\ref{fig:Distribution-indegree}).  
This implies the existence of influential accounts in both pro- and anti-vaccine groups, which is often the case in a spreading phenomenon.

Furthermore, we compared the network characteristics of these two communities that changed after mass vaccination (i.e. December 2020). As shown in Table \ref{tab:network-analysis-time}, overall, the difference between two period is minute; only changes in network density significantly decreased. When it comes to two groups, the sizes (nodes and links) of pro-vaxxers increased but network density decreased in the after period. Conversely, those of anti-vaxxers decreased, but network density increased in the after period. We can infer from these results that although the pro-vaccine community grew once mass vaccination began, they became sparser than before. However, although the anti-vaccine community is comparatively small and lose members once the vaccination started, those became more tightly knit.

\begin{table*}[h]
\centering
\caption{Network features of the pro- and anti-vax clusters in the retweet network.}
\label{tab:network-analysis}
\begin{tabular}{cclclcl}
\toprule
Measure                & \multicolumn{2}{c}{Overall}  & \multicolumn{2}{c}{Pro-vaxxers} & \multicolumn{2}{c}{Anti-vaxxers} \\ \midrule
Nodes (\# users)                  & \multicolumn{2}{c}{77,934}    & \multicolumn{2}{c}{11,522}       & \multicolumn{2}{c}{6,909}         \\
Links (\# unique RTs)                 & \multicolumn{2}{c}{1,999,164}  & \multicolumn{2}{c}{207,343}      & \multicolumn{2}{c}{171,589}       \\
Network Density        & \multicolumn{2}{c}{3.29E-04} & \multicolumn{2}{c}{1.56E-03}    & \multicolumn{2}{c}{3.60E-03}     \\
Clustering Coefficient & \multicolumn{2}{c}{0.039}         & \multicolumn{2}{c}{0.045}       & \multicolumn{2}{c}{0.070}        \\
Average Distance       & \multicolumn{2}{c}{6.696}         & \multicolumn{2}{c}{1.581}       & \multicolumn{2}{c}{1.393}        \\
\bottomrule
\end{tabular}
\end{table*}

 \begin{table*}[]
 \centering
 \caption{Network features before and after mass vaccination.}
 \label{tab:network-analysis-time}
 \begin{tabular}{@{}cclclcl@{}}
 \toprule
 Measure & \multicolumn{2}{c}{Overall}            & \multicolumn{2}{c}{Pro-vaxxers}       & \multicolumn{2}{c}{Anti-vaxxers}     \\ \midrule
  & Before & After & Before & After & Before & After \\ \hline
 Nodes (\# users)    & 34,283  &  36,441  & 4,877  & 7,031   & 3,763  &  3,280  \\
 Links (\# unique RTs)   & 736,888 &  733,073 & 60,258 & 100,897 & 75,950 & 63,492 \\
 Network Density & 6.27E-04 & 5.52E-04 & 2.53E-03 & 2.04E-03& 5.37E-03 & 5.90E-03 \\ \bottomrule
 \end{tabular}
 \end{table*}

We found that among the top 10 users in these clusters, six accounts belonging to anti-vaxxers have been banned by Twitter, while none of the pro-vaxxer accounts have been banned. 
This observation indicates that our classification of pro- and anti-vaxxers is correct and reliable.

\subsection{Psycho-linguistic features of pro- and anti-vaxxers}
After identifying pro- and anti-vaccine communities, we compared the two in terms of psycho-linguistic features, including emotion. 
Specifically, we selected categories such as analytical thinking, affective processes, and personal concerns from LIWC.
The LIWC categories and subcategories we used are shown in Table \ref{tab:LIWC2015-information}. 
In the following, we used the independent $t$-test to compare the average score between anti- and pro-vaxxers. 
For all tests, the confidence level is 95\%. The confidence intervals were computed from the bootstrapped samples (We randomly sampled tweets ($n$=45,809) and replies ($n$=254,660) and repeated it multiple times ($n$=10,000) for statistical evaluation.) Additionally, to find out the differences between tweets and replies, we separately listed the scores of the two communities.

\begin{figure*}[!t]
    \centering
    \includegraphics[width=\linewidth]{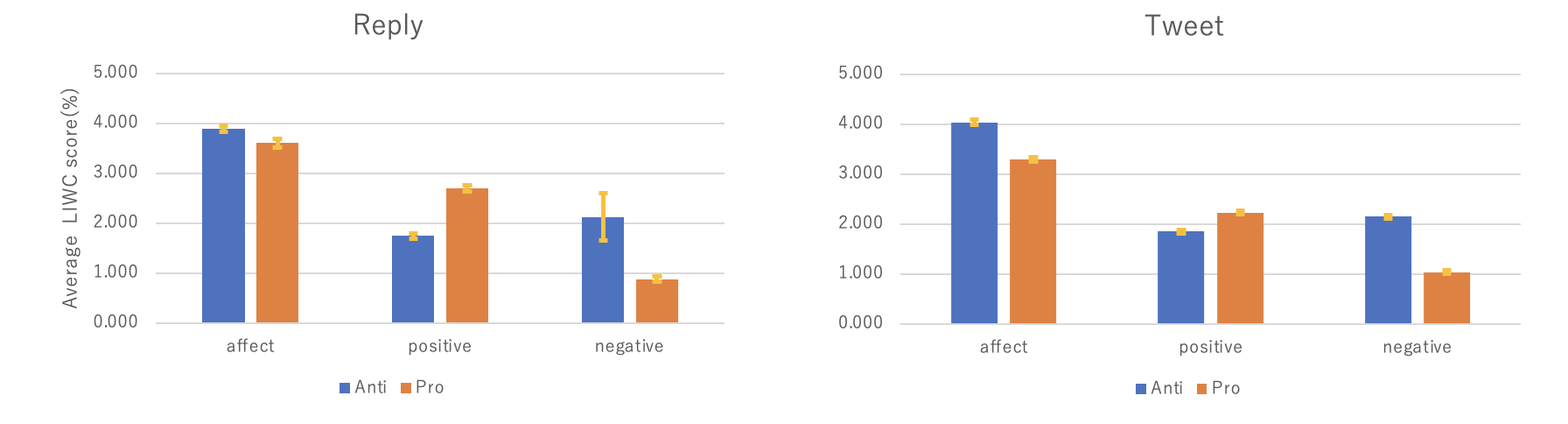}
    \caption{Average emotion score (\%) for anti- and pro-vaxxers. Affective processes including positive and negative emotion.Differences between anti- and pro-groups are all significant (independent $t$-test, $p<0.001$)}
    \label{fig:eomtion-liwc}
\end{figure*}

Affective processes include several emotional words in LIWC. Examples of positive emotion words are `love', `nice', and `sweet', while words such as `hurt', `ugly', and `nasty' are seen as negative. Figure \ref{fig:eomtion-liwc} shows that, overall, anti-vaxxers expressed more affect (positive and negative combined) than pro-vaxxers in both tweets and replies. If we consider positive and negative affect separately, we find that in tweets and replies, anti-vaxxers have higher negative and lower positive emotions, while the opposite is true for pro-vaxxers.

\begin{figure*}
    \centering
    \includegraphics[width=\linewidth]{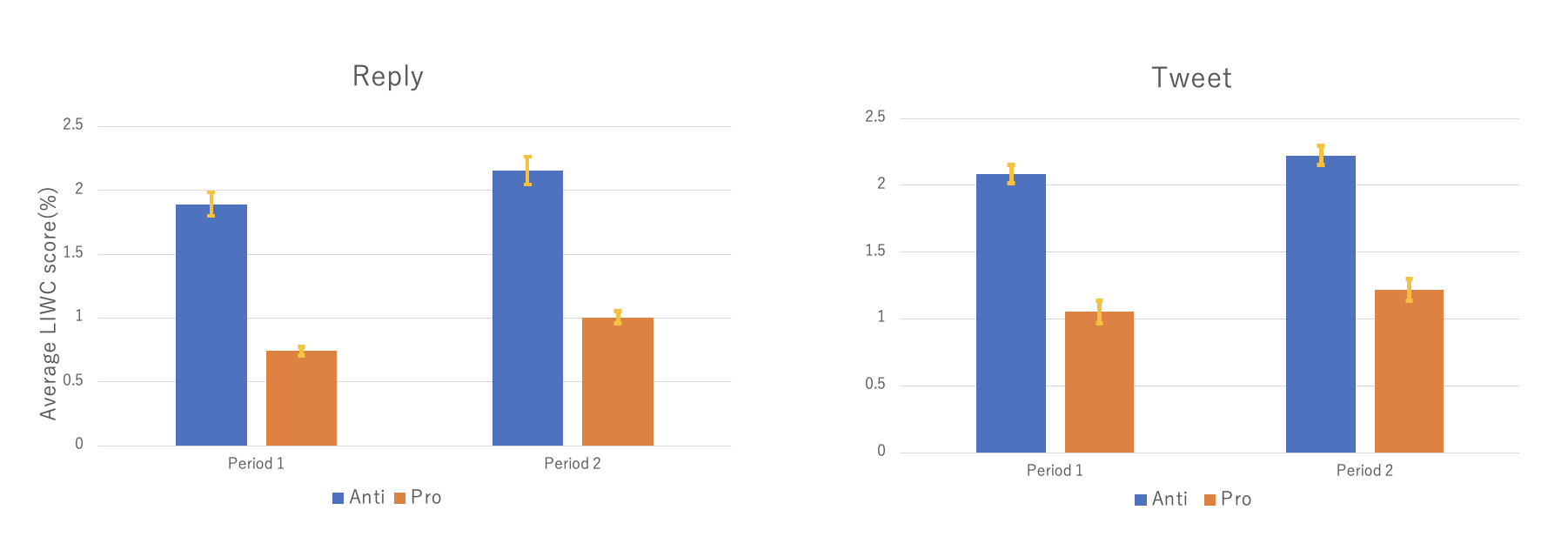}
    \caption{Average negative emotion scores before and after the mass vaccination (Period 1: February 2020 to November 2020; Period 2: December 2020 to June 2021). Differences between anti- and pro-groups are all significant (independent $t$-test, $p < 0.001$).}
    \label{fig:eomtion-liwc-time}
\end{figure*}

Mass vaccination for COVID-19, which was expected to improve negative emaitions both in pro- and anti-vaccination groups, started in December 2021 \cite{owidcoronavirus}.
However, Figure~\ref{fig:eomtion-liwc-time} shows that negative emotions instead increased before the mass vaccination (Period 1: February 2020 to November 2020) and after it (Period 2: December 2020 to June 2021). In addition, for replies and tweets, anti-vaxxers showed higher negative emotions compared to pro-vaxxers during Period 1. This trend increased in Period 2, and similarly anti-vaxxers' expression of emotion was larger than that of pro-vaxxers. These results suggest that anti-vaxxers propagated their anti-vaccination beliefs more passionately and emotionally even after the start of mass vaccination~\cite{guo2019exploiting}. 

\begin{figure*}[h]
    \centering
    \includegraphics[width=\linewidth]{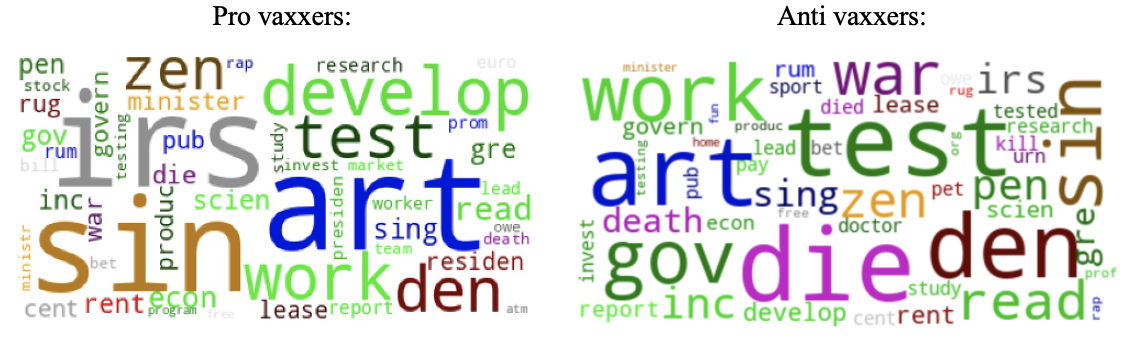} 
    \caption{Word clouds for pro- and anti-vaxxers. Colours correspond to different LIWC subcategories: green for work, blue for leisure, red for home, black for money, orange for religion, and purple for death. }
    \label{fig:wordcloud}
\end{figure*}

There are subcategories of personal concerns in LIWC. In Fig.~\ref{fig:wordcloud}, the frequency of words used in personal concerns for anti- and pro-vaxxers are visualized using Wordcloud. We found that the three most highly used subcategories by pro-vaxxers are money (black), religion (orange), and leisure (blue), while anti-vaxxers have shown a higher usage of death (purple) and work (green). 

\begin{figure*}[ht]
    \centering
    \includegraphics[width=\linewidth]{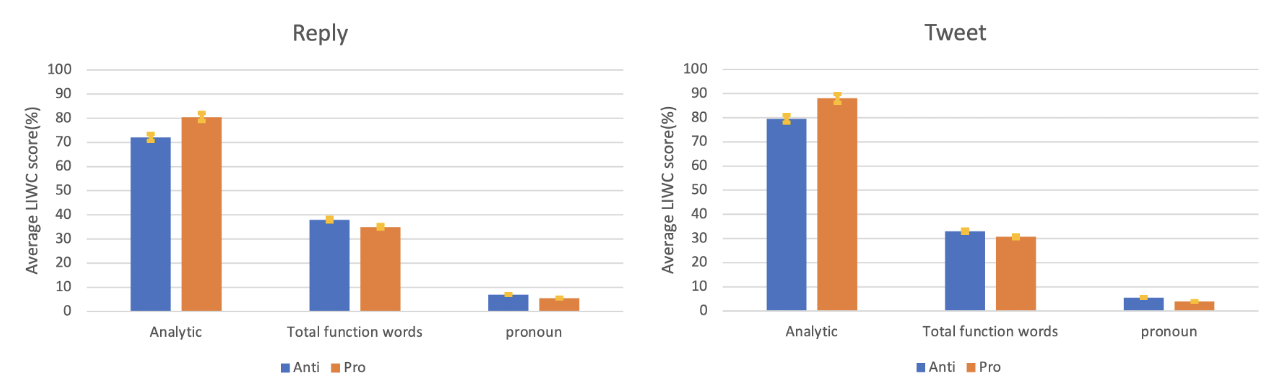}
    \caption{LIWC scores for analytical, total function words, and total pronouns on average for anti- and pro-vaxxers. Differences between anti- and pro-groups are all significant (independent $t$-test, $p < 0.001$).}
    \label{fig:narrative-liwc}
\end{figure*}

\begin{figure}[h]
    \centering
    \includegraphics[width=\linewidth]{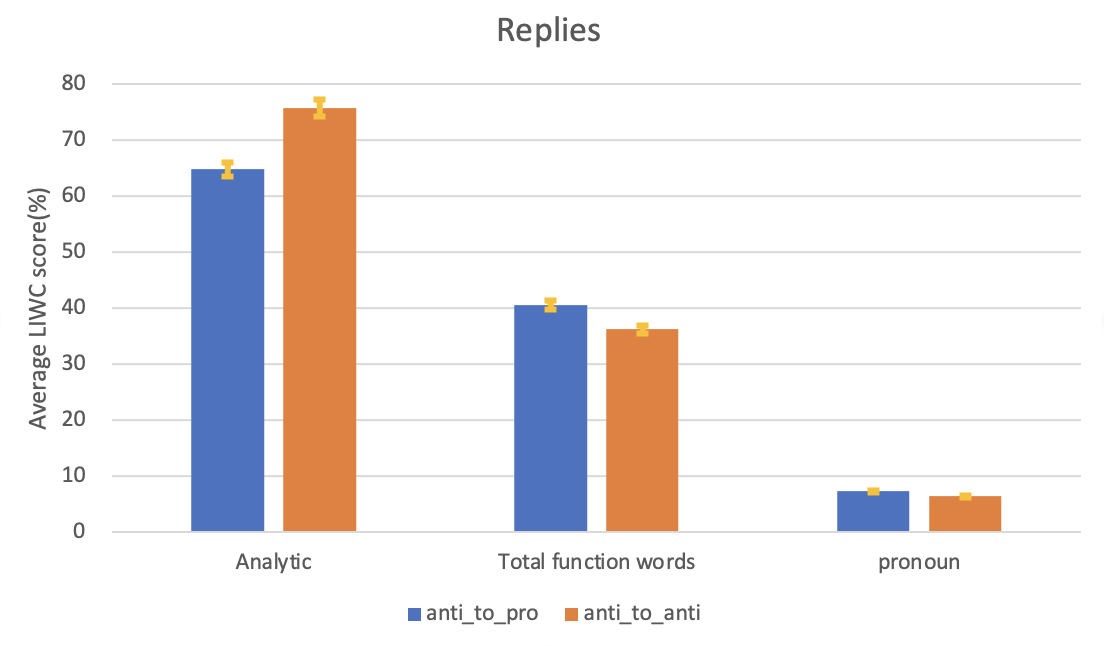} 
    \caption{LIWC scores for analytical, total function words, and total pronouns in replies between anti-vaxxers and between anti- and pro-vaxxers.  Differences are all significant (independent $t$-test, $p < 0.001$).}
    \label{fig:within_between_analytic}
\end{figure}

Comprehending the thought process (whether analytical or narrative thinking) of both pro- and anti-vaxxers can also provide critical information that can be utilized to create a mitigation strategy. In both replies and tweets, we found that anti-vaxxers used more function words and pronouns, and had a lower analytic score than pro-vaxxers. Recall that the higher function words, pronouns score and lower analytic scores represent narrative thinking. This suggests that anti-vaxxers use more narrative thinking than pro-vaxxers (Fig.~\ref{fig:narrative-liwc}). Similarly, we found that replies showed higher usage of function words, pronouns and a lower analytic score than tweets. This result also indicates that the trend of narrative thinking can be stronger in replies, a targeted message).
Furthermore, we compared analytic thinking tendency in replies between anti-vaxxers and between anti- and pro-vaxxers. It turns out that anti-vaxxers were in the narrative mode when replying to pro-vaxxers (Fig.~\ref{fig:within_between_analytic}). 

\subsection{Moral foundations in pro- and anti-vaxxers}
Morality is another psycho-linguistic feature. It is important to explain the process of making social judgements \cite{singh2021morality}. As explained, we used MFD to assess morality tendencies among pro- and anti-vaxxers. The comparison of the average scores of both groups is shown in Table \ref{tab:ttest-mfd}. 
We can see that anti-vaxxers show higher scores in the Vice moral foundations (e.g. Harm, Cheating, Betrayal, Subversion, Degradation), indicating they tend to use moral violating words. 
While in the Virtue moral foundations, pro-vaxxers show higher scores in Care, Fairness, Loyalty and Sanctity, indicating they expressed moral content. 
Taken together, anti-vaxxers frequently used immoral language in their posts, thereby distributing anxiety-provoking (fake)news and messages.

\begin{table}[t]
\caption{Comparing the average score of Pro-vaxxers (Pro) and Anti-vaxxers (Anti). $p$-values from independent $t$-test are listed.}
\label{tab:ttest-mfd}
\centering
\begin{tabular}{ccc}
\toprule
Measure     & Mean Diff        & $p$-value                \\ \midrule
Care        & Pro \textgreater{}Anti   & \textless0.01   \\
Harm        & Pro \textless{}Anti & \textless0.0001\\ \midrule
Fairness    & Pro \textgreater{}Anti & \textless0.0001 \\
Cheating    & Pro \textless{}Anti & \textless0.0001 \\ \midrule
Loyalty     & Pro \textgreater{}Anti & \textless0.0001 \\
Betrayal    & Pro \textless{}Anti & \textless0.0001 \\ \midrule
Authority   & n.s.                         & n.s. \\
Subversion  & Pro \textless{}Anti & \textless0.0001 \\ \midrule
Sanctity    & Pro \textgreater{}Anti & \textless0.0001 \\
Degradation & Pro \textless{}Anti & \textless0.0001 \\
\bottomrule
\end{tabular}
\end{table}

\section{Discussion}
We have shown that anti-vaxxers used more affective process and negative words, while pro-vaxxers used more positive words (Fig.~\ref{fig:eomtion-liwc}).  Another research also showed that the strategy of the anti-vax groups involves strong emotions with highly toxic and negative words during the early COVID-19 pandemic~\cite{miyazaki2021strategy,Miyazaki2022-tz}. Thus, our finding is consistent with the previous result.
In addition, we showed that anti-vaxxers tend to use vice moral languages in all of the five moral foundations, whereas pro-vaxxers exhibit the opposite tendency.
These findings are the answer to RQ1.

We have seen positive effects globally since the mass vaccination started. How did anti-vax communities change their interactions on Twitter after the mass vaccination? Our results showed that, compared to pro-vaxxers, anti-vaxxers expressed even higher negative emotions after the mass vaccination, which suggests their firmness (Fig. \ref{fig:eomtion-liwc-time}).
These results are the answer to RQ2.
Strangely, similar results were observed in pro-vaxxers, although we expected that pro-vaxxers became more positive emotionally. 
These results cannot be explained by the `backfire effect', where exposure to opposing views can lead to the increased commitment to preexisting beliefs \cite{nyhan2010corrections}.
Therefore, further investigation is needed to unveil this emotional shift phenomenon.

In addition, we found that anti-vaxxers show more narrative thinking (higher score of total function words and total pronouns) and a lower score in analytic thinking, as shown in Fig. \ref{fig:narrative-liwc}. Unlike analytic thinking, narrative thinking tends to be more personal, and rumours are easier to spread \cite{singh2017automated}. However, a previous study showed that informed groups (pro-vaxxers) have more narrative thinking \cite{memon20201characterizing}, which is contrary to our results but that study was conducted on a small dataset for a short period (less than a month). We used larger and more longitudinal data, and our results show that narrative thinking is more dominant among anti-vaxxers, especially during the COVID-19 pandemic. As mentioned above, anti-vaxxers showed negative moral tendencies (i.e., usage of vice words) based on MFD, consistent with anti-vaxxers using more negative emotions because vice words are correlated with those \cite{matsumoto2005algorithm}. Fig. \ref{fig:wordcloud} shows that anti-vaxxers use more `death' words, a subcategory belonging to personal concern. This result suggests that anti-vaxxers may be more concerned about death caused by vaccination.
 
Because replying can reach beyond follow-follower relations, reply patterns on Twitter can also provide valuable information, and previous studies have provided such evidence (e.g., \cite{usher2018twitter}, \cite{bliss2012twitter}, \cite{nishi2016reply}). We also investigated reply patterns to understand the psycho-linguistic features of Twitter conversations on vaccination. We found that replies showed more narrative thinking (Fig. \ref{fig:narrative-liwc}), because replying is an immediate reaction and thus less inclined toward analytic thinking.

These findings provide useful implications for spreading credible content from pro-vaxxers while reducing the exposure of misinformation and anxiety-provoking posts from ant-vaxxers. 
Because anti-vaxxers' posts are typically charged with higher negativity and vice morality, algorithms can detect their harmful content by measuring psycho-linguistic features. 
Then, the SNS system can hide such posts until users agree to avoid unnecessary exposure.
At the same time, psycho-linguistic features as well as network structures help us identify those pro-vaxxers who spontaneously transmit trustful vaccine information in near real-time. 
Thus, we may leverage such information for social fact-checking to oppose anti-vaccine narratives at scale.

\section{Limitations and Future Work}
Several previous studies have shown that Twitter and other social media platforms are not representative of the general population \cite{mislove2011understanding,greenwood2016social,mellon2017twitter}. Even though our findings are statistically significant, we should be aware of the gap between online social networks and reality. Besides this, the focus of our study is only Twitter, but there are several other popular social media platforms such as Facebook, Reddit, and Instagram. Analysis of vaccination communities on these platforms can also bring forward critical insights. In our future study, we will focus on the comparative analysis of vaccination communities on Twitter, Facebook, Instagram, and Reddit.
Comparison between different languages and cultures is also an important future direction (e.g. \cite{Miyazaki2022-tz}). 

\section{Conclusion}
In this study, we quantified psycho-linguistic differences among competing vaccination communities on Twitter during the COVID-19 pandemic. Based on the differences in linguistic usage, we found that anti-vaxxers tend to show more negative, narrative thinking, and immoral tendencies. In terms of network difference, anti-vaxxers showed a tighter network structure and strengthening of their anti-vaccine beliefs. 

The mass vaccination of people during the COVID-19 pandemic shows that these vaccines are working. However, even this news does not deter anti-vaxxers from their beliefs; rather; it strengthens these negative emotions. This vicious circle needs to break if we want to immunize all high-risk people in the world and achieve herd immunity, while preventing the online anti-vaccine movement. Our results provide key insights for developing countermeasures against the online anti-vaccine movement, both at individual and society levels. 

\section*{Acknowledgement}
We would like thank the members of CREST projects (JPMJCR20D3 and JPMJCR17A4) for fruitful discussions.

\section*{Financial Support}
This work was supported by JST, CREST Grant Number JPMJCR20D3, Japan.


\section*{Statement of interest}
None.

\vskip2pc

\vskip2pc

\noindent \large \textbf{Biographies}

\vskip2pc

\noindent\normalsize\textbf{Jialiang Shi} received the B.E. degrees from the Chang'an University, Xi'an, China in 2019. He is currently enrolled as a Master student at Graduate School of Infomatics, Nagoya University. His research interests focus on fake news on social media and natural language processing.

\vskip2pc

\noindent\textbf{Piyush Ghasiya} received Ph.D. from Kyushu University in 2020. Presently, he is a postdoctoral research fellow at School of environment and Society, Tokyo Institute of Technology. His research interests include computational social science and international relations. 

\vskip2pc

\noindent\textbf\noindent\textbf{Kazutoshi Sasahara} received his Ph.D. from The University of Tokyo in 2005. He worked as a researcher at several institutions including RIKEN, The University of Tokyo, and UCLA. From 2012 to 2020, he was an Assistant Professor in Graduate School of Informatics, Nagoya University. Since 2020, he is an Associate Professor in School of Environment and Society, Tokyo Institute of Technology. His research interests are computational social science and social innovation.

\vskip2pc


\vskip1pc



\end{document}